\documentclass[proceedings]{JHEP}

\title{Testing the AdS/CFT correspondence beyond large N}

\author{ Adel Bilal and Chong-Sun Chu  \\  
 Institute of Physics,
University of Neuch\^atel, CH-2000 Neuch\^atel, Switzerland\\
E-mail: \email{Adel.Bilal@iph.unine.ch }, 
\email{chong-sun.chu@iph.unine.ch}  
}

\conference{Quantum Aspects of Gauge Theories, Supersymmetry and
Unification}  

\abstract{
According to the AdS/CFT correspondence, the $d=4$, ${\cal N}=4$ 
${\rm SU}(N)$ super Yang-Mills theory is dual to the type IIB 
string theory compactified on $AdS_5 \times S^5$. Most of the tests 
performed so far are confined to the leading order at large $N$ or
equivalently string tree-level. To probe the correspondence beyond 
this leading order and obtain $1\over N^2$ corrections is difficult 
since string one-loop computations on an $AdS_5\times S^5$ background 
generally are beyond feasibility. However, we will show that the 
chiral ${\rm SU}(4)_R$ anomaly of the super YM theory provides an 
ideal testing ground to go beyond leading order in $N$. In this 
paper, we review and develop further our previous results [1] 
that the $1/N^2$ corrections to the chiral anomaly on the super YM 
side can be exactly accounted for by the supergravity/string effective
action induced at one loop.
}

\def\cN{{\cal N}}

\newfont{\goth}{eufm10 scaled \magstep1}

\def\a{\alpha}

\def\c{\gamma}
\def\d{\delta}
\def\e{\epsilon}

\def\k{\kappa}
\def\l{\lambda}
\def\m{\mu}
\def\n{\nu}

\def\r{\rho}
\def\s{\sigma}

\def\beq{\begin{equation}}\def\eeq{\end{equation}}
\def\beqa{\begin{eqnarray}}\def\eeqa{\end{eqnarray}}
\def\barr{\begin{array}}\def\earr{\end{array}}

\def \Ds {{D \hspace{-7.4pt} \slash}\;}

\def \ys {{y\kern-.5em / \kern.3em}}
\def\ads{$AdS_5 \times S^5$ \,}
\def\R{{\bf R}}
\def\un{{\rm U}(N)}
\def\ap{\alpha'}
\def\Tr{{\rm Tr}\, }

\let\bm=\bibitem

\def\nn{\nonumber}
\def\bd{\begin{document}}
\def\ed{\end{document}}
\def\ba{\begin{array}}
\def\ea{\end{array}}
\def\bea{\begin{eqnarray}}
\def\eea{\end{eqnarray}}
\def\ft#1#2{{\textstyle{{\scriptstyle #1}\over {\scriptstyle #2}}}}
\def\fft#1#2{{#1 \over #2}}
\newcommand{\be}{\begin{equation}}
\newcommand{\ee}{\end{equation}}
\newcommand{\eq}[1]{(\ref{#1})}
\def\eqs#1#2{(\ref{#1}-\ref{#2})}
\def\det{{\rm det\,}}
\def\tr{{\rm tr}}
\newcommand{\ho}[1]{$\, ^{#1}$}
\newcommand{\hoch}[1]{$\, ^{#1}$}
\def\ra{\rightarrow}
\def\uha{{\hat {\underline{\a}} }}
\def\uhc{{\hat {\underline{\c}} }}

\begin{document}

\section{Introduction}

We begin by briefly reviewing some relevant basic aspects of the 
AdS/CFT correspondence \cite{c1,c2,c3}, see in particular \cite{rev}.

Consider type IIB string theory with a number $N$ of D3 branes. 
There are open strings ending on these D3 branes and closed 
strings in the bulk. The effective low energy action consists 
of IIB supergravity describing the bulk closed strings, and 
$\cN=4$ supersymmetric $\un$ gauge theory (SYM) describing the 
open strings ending on the D3 branes. The gauge group of the latter 
actually is ${\rm SU}(N)$ since the ${\rm U}(1)$ decouples. 
Considering low energies at fixed $\ap$ is equivalent to fixed energy 
and taking the $\ap\to 0$ limit. In this limit the gravitational 
coupling $\k\sim g_s \ap^2$ vanishes and the interactions between the 
branes and the bulk can be neglected, as well as all higher 
derivative terms in the brane action. Only free bulk supergravity 
and pure $\cN=4$ ${\rm SU}(N)$ SYM in $d=4$ dimensions remain. The 
latter is a conformal field theory (CFT).

There is a different way to describe the same physics. D3 branes 
may be viewed as certain solutions of the supergravity field 
equations, namely
\bea\label{ui}
&& 
{\rm d}s^2={1\over \sqrt{f}} 
\left( -{\rm d}t^2 + {\rm d}x_1^2 + {\rm d}x_2^2 
+ {\rm d}x_3^2 \right)
\nn\\
&&\quad\quad
+ \sqrt{f} \left( {\rm d}r^2 + r^2 {\rm d}\Omega_5^2 \right), 
\nn\\
&& f=1+{R^4\over r^4} \ , \quad R^4=4\pi g_s \ap^2 N  \ .
\eea
Here $t,x_1,x_2,x_3$ are the longitudinal coordinates (on the D3)
while $r$ and $\Omega_5$ describe the transverse space. $N$ is 
the number of (coincident) D3 branes. There is also a five-form field 
strength $F$ which is proportional to $N$. Again, one wants to study 
the low energy excitations in this description and compare with the 
first one. Due to the non-trivial function $f(r)$ in the metric 
there is a red-shift factor between energies measured at $r$ and 
energies measured at $r=\infty$:
\be\label{uii}
E_\infty= f^{-1/4} E_r 
= \left( 1+{R^4\over r^4}\right)^{-1/4} E_r \ .
\ee
One sees that if $r\to 0$, $E_\infty\to 0$ for any finite $E_r$. 
So there are two types of low energy excitations: low energy at 
finite $r$ (bulk) or finite energy near the horizon ($r=0$), and 
the two types decouple yielding (free) bulk supergravity and near 
horizon supergravity. Concentrate on the near horizon limit: 
as $r\to 0$ one has $f\sim {R^4\over r^4}$. Since also 
$\ap\to 0$, it is convenient to introduce the finite quantity 
$u=r/\ap$ so that the near horizon metric becomes 
\bea\label{uiv}
&& {1\over \ap} {\rm d}s^2 = \l^{-1/2} u^2
\left( -{\rm d}t^2 + {\rm d}x_1^2 + {\rm d}x_2^2 
+ {\rm d}x_3^2 \right)
\nn\\
&&\quad\quad\quad + \l^{1/2} {{\rm d}u^2\over u^2}
+ \l^{1/2} {\rm d}\Omega_5^2 \ ,
\eea
where we introduced the finite quantity
\be\label{uv}
\l=4\pi g_s N={R^4\over \ap^2} \ .
\ee
The metric \eq{uiv} is the metric of $AdS_5\times S^5$.

Comparing both descriptions of the same phy\-sics one then is led to 
identify the conformally invariant, $\cN=4$ ${\rm SU}(N)$ SYM theory 
in $d=4$ with the supergravity or small $\ap$ limit of IIB string 
theory on $AdS_5\times S^5$.  We will be more precise shortly. 

How do the different parameters on the SYM side compare to those 
of the supergravity/string theory? In the former we have the 
coupling $g_{\rm YM}$ and the integer $N$ determining the gauge group
${\rm SU}(N)$. On the
supergravity/string side we have $\ap$ and $R$ (with only the 
dimensionless ratio $R^2/\ap$ being a relevant parameter) and 
the coupling $g_s$. A first relation is already obtained in eq. 
\eq{ui}, namely $R^4=4\pi g_s \ap^2 N$ or equivalently eq. 
\eq{uv}. A second relation is obtained from the D3 brane action 
from which one reads the YM coupling in terms of the string 
coupling. Thus
\be\label{uvi}
{1\over g_{\rm YM}^2} = {1\over 4\pi\, g_s} \quad {\rm and} \quad
N= {1\over 4\pi\, g_s} {R^4\over \ap^2}
\ee
expresses the SYM parameters $g_{\rm YM}$ and $N$ in terms of 
the string/supergravity parameters $g_s$ and $R^2/\ap$ and vice 
versa. In large $N$ gauge theories the relevant loop-counting 
parameter is the 't Hooft coupling $g_{\rm YM}^2 N$ rather than 
$g_{\rm YM}^2$. Combining both eqs \eq{uvi} yields 
$g_{\rm YM}^2 N={R^4\over \ap^2}$ which by eq \eq{uv} is just 
the quantity called $\l$. It is useful to rewrite the relations 
between the parameters of the two descriptions as
\be\label{uvii}
\l={R^4\over \ap^2} \quad , \quad {N\over \l}={1\over 4\pi \, g_s}
\ee
with $\l$ now meaning the 't Hooft coupling.

Let us first comment on the first relation: perturbative SYM 
theory is a good description if the 't Hooft coupling $\l$ is 
small. Supergravity, rather than string 
theory, should be a good description if the radius of curvature 
of $AdS_5$ and $S^5$ is large, meaning $R^2\gg \ap$ or $\l$ 
large. The two regimes are opposite as is often the case with 
dualities. This of course avoids the obvious contradiction that 
both descriptions look so different.

At fixed 't Hooft coupling $\l$, the second relation \eq{uvii} 
tells us that ${1\over N^2}$ corresponds to the string 
loop-counting parameter $g_s^2$, so that the large $N$ limit of 
SYM corresponds to classical string theory (or classical 
supergravity if also $\l\gg 1$), and ${1\over N^2}$ corrections 
should correspond to one-loop effects in string theory.

One now has various possible conjectures:
1.) The weakest one is: SYM is dual to $AdS$ supergravity only 
for $\l\to\infty$, but the full string theory is different. This 
version would not be very useful. 2.) The SYM theory is dual to 
string theory on $AdS$ for finite $\l$ but only as $N\to\infty$ 
or equivalently $g_s\to 0$. This includes $\ap$ corrections 
beyond the supergravity approximation, but no string loops. 
3.) This is the strongest version, generally referred to as the 
Maldacena conjecture: the SYM theory is dual to string theory 
for all $\l$ and all $N$, i.e. all $R^4/\ap^2$ and $g_s$.

While there is now reasonable evidence for version 2.) of the 
conjecture (see e.g. \cite{rev}) the strong version 3.) is hard to 
test since standard string or supergravity loop computations on an 
$AdS_5\times S^5$ background are difficult, to say the least, if 
not unfeasible, with the present state of the art.

Many successful tests are group theoretic in nature. Some are 
not restricted to tree-level or even perturbation theory, but on 
the other hand they do not really provide any real test at 
one-loop. Examples are global symmetries (disregarding possible 
anomalies for the moment): $AdS_5$ space-time has an 
${\rm SO}(4,2)$ symmetry which also is the conformal group of the
$\cN=4$ SYM theory. The ``internal" symmetry is 
${\rm SO}(6)\simeq {\rm SU}(4)$: this is the isometry of the 
$S^5$ as well as the R-symmetry of the SYM theory. The latter 
actually is anomalous which will be important for us. Both theories 
have the same amount of supersymmetry, the full supergroup being 
${\rm SU}(2,2\vert 4) \supset$ ${\rm SO}(4,2) \times {\rm SU}(4)_R$. 
Also the duality symmetry ${\rm SL}(2, {\bf Z})$ is the same as 
it acts on 
\be\label{uviii}
\tau={4\pi i\over g^2_{\rm YM}} + {\theta\over 2\pi} 
= {i\over g_s}+{\chi\over 2\pi} \ .
\ee

A promising arena for performing tests beyond the large $N$ limit 
or string tree-level is to look at certain anomalies. On the SYM 
side anomaly coefficients are easily established one-loop effects 
in $\l$ that are protected against higher order corrections. 
Typically such an anomaly coefficient will depend on the number of 
fermion fields in the SYM theory, i.e. on $N$. The goal then is to 
reproduce the exact $N$-dependence from the string theory 
including subleading terms  $\sim {1\over N^2}$ coming from string 
loops. If 
we want to have any chance to be able to do this calculation, the 
relevant quantity to compute in string theory should be of a 
topological nature, like a Chern-Simons term, so that the actual 
metric on $AdS_5$ is irrelevant.

\section{The chiral ${\rm SU}(4)_R$ anomaly in the $\cN=4$ SYM}

The anomaly we will consider is the chiral ${\rm SU}(4)_R$
anomaly. As already mentioned, ${\rm SU}(4)_R$ is a (classical) 
global symmetry of the $\cN=4$ SYM theory. Due to the presence of 
chiral fermions transforming in complex conjugate representations 
of ${\rm SU}(4)_R$ this symmetry is spoiled at one loop and there 
is an anomaly: the one-loop effective action in the presence of 
{\it external} ${\rm SU}(4)_R$ gauge fields is no longer invariant under 
${\rm SU}(4)_R$ and the non-invariance is proportional to the 
number of fermions. Since they are also in the adjoint 
representation of the ``true" gauge group ${\rm SU}(N)$ there are 
$N^2-1$ of them, and the anomaly is proportional to $N^2-1$. As we 
recall below, the leading term $\sim N^2$ is accounted for by 
tree-level supergravity \cite{dzf}. It is the $-1$ correction 
which should originate from a string/supergravity loop effect, and 
it indeed does as we showed in \cite{bc} and explain in the 
remainder of this paper.

Before explaining the string/supergravity loop correction let us 
review how the leading $N^2$ term is obtained in the 
string/supergravity description. Here 
${\rm SU}(4)\simeq {\rm SO}(6)$ acts as an isometry on the $S^5$. 
As a consequence, the $AdS_5$ supergravity is actually a gauged 
supergravity \cite{GRW1, GRW2, vN1} and there is an ${\rm SU}(4)_R$ 
gauge group with 
gauge fields $\tilde A_\m^a(x,z)$, $\m=0,1,\ldots 4$ and 
$a=1, \ldots 15={\rm dim}\, {\rm SU}(4)$. This gauge group is of 
course not to be confused with the ${\rm SU}(N)$ of the conformal 
SYM theory. Note also that in the latter, ${\rm SU}(4)_R$ is a 
global symmetry, hence there are ${\rm SU}(4)_R$ currents 
$J_\m^a(x)$, $\m=0,1,2,3$, but no associated gauge fields. We can 
nevertheless couple these currents to {\it external} gauge fields 
$A_\m^a(x)$, $\m=0,1,2,3$ which act as sources for these currents. 
Then by a standard argument, the non-invariance of the  
one-loop effective action $\Gamma[A_\m]$ under gauge 
transformations of these external gauge fields is equivalent to 
the covariant non-conservation of the currents: let $\delta_v$ be 
such a gauge transformation with parameter $v$, then
\bea\label{uix}
&&\delta_v \Gamma[A_\m]
=\int \delta_v A_\m^a {\delta\Gamma\over \delta A_\m^a} 
= \int \delta_v A_\m^a J^{a,\m}
\nn\\
&&
=\int (D_\m v)^a  J^{a,\m}
= - \int v^a (D_\m J^\m)^a
\eea
with 
\be\label{ux}
(D_\m J^\m)^a \sim -(N^2-1) d^{abc} \epsilon^{\m\n\r\s}
\partial_\m A_\n^b \partial_\r A^c_\s + \ldots
\ee
the precise numerical coefficient being given below.

There is a standard prescription \cite{c3} in the AdS/CFT 
correspondence how to compute correlation functions: we will give 
this prescription for the case of present interest. For any 
(${\rm SU}(N)$ gauge-invariant) operator ${\cal O}(x)$ like the 
currents $J_\m^a(x)$ of the SYM theory, introduce a source 
$\phi_0(x)$ like the $A_\m^a(x)$. Then the generating functional 
for correlators of $J_\m^a$ is just
\be\label{uxi}
{\rm e}^{-\Gamma[A]}
\equiv \langle 
{\rm e}^{-\int{\rm d}^4x A_\m^a(x) J^{a,\m}(x)} \rangle_{\rm SYM} \ .
\ee
In $AdS_5$ string theory there is a field $\phi(x,z)$ such that at 
the boundary $z=0$ of $AdS_5$ (note that $z=1/u$) which is just 
the four-dimensional space of the SYM theory one has 
$\phi(x,z=0)=\phi_0(x)$. In our case this is just 
$\tilde A_\m^a(x,z=0)=A_\m^a(x)$ (for $\m=0,1,2,3$ only) where the 
$\tilde A_\m^a$ are the gauge fields of the gauged supergravity. 
The prescription \cite{c3} then is
\be\label{uxii}
{\rm e}^{-\Gamma[A]}
=Z_{\rm string} \Big\vert_{\tilde A_\m^a(x,z=0)=A_\m^a(x)} \ ,
\ee
meaning that the string partition function should be evaluated 
subject to the boundary condition $\tilde A_\m^a(x,z=0)=A_\m^a(x)$ 
for $\m=0,1,2,3$. Writing
\be\label{uxiii}
Z_{\rm string}={\rm e}^{-S^{\rm cl}_{\rm string} 
-S^{\rm 1-loop}_{\rm string}-\ldots}
\equiv {\rm e}^{-S_{\rm string}^{\rm eff}}
\ee
eq \eq{uxii} together with eq \eq{uix} implies that, if the 
AdS/CFT correspondence is correct, we should have
\bea\label{uxiv}
&&\delta_v S_{\rm string}^{\rm eff} 
\Big\vert_{\tilde A_\m^a(x,z=0)=A_\m^a(x)}
=\delta_v  \Gamma[A] 
\nn\\
&&\qquad\qquad
= - \int v^a (D_\m J^\m)^a
\eea
which is non-vanishing according to \eq{ux}. Thus the 
${\rm SU}(4)_R$ gauge variation of $S_{\rm string}^{\rm eff}$ 
should directly reproduce the SYM chiral ${\rm SU}(4)_R$ anomaly. 
Actually for the purpose of reproducing the  leading $N^2$ part of 
the anomaly it is enough to consider the classical supergravity 
action \cite{dzf}.

Let us now determine the exact anomaly coefficient of the $\cN=4$
SYM theory in 4 dimensions. This theory has
four complex
Weyl fermions $\l$ in  the fundamental representation of 
${\rm SU}(4)_R$
with the chirality part (0,1/2) in ${\bf 4}$ and (1/2,0) 
in ${\bf 4}^*$ (see for example \cite{FFZ}. 
Our conventions here  are equivalent to those of \cite{FFZ}.) 
Moreover, all fields are also in the adjoint representation of the 
``true" gauge group ${\rm SU}(N)$ which acts as a ``flavour" group 
with  respect to the ${\rm SU}(4)_R$. Thus there are actually 
$N^2-1$ complex Weyl fermions $\l$ in the ${\bf 4}$, resp. 
${\bf 4}^*$.
The correctly normalised R-symmetry anomaly is given by
\be \label{YManom}
\delta_v \Gamma[A] =  (N^2-1) \; \int_{S^4} \omega^1_4(v,A).
\ee
The differential forms
\bea
\omega^1_4 (v,A) = \frac{1}{24 \pi^2} \Tr [v d(AdA +\frac{1}{2}A^3)],\nn\\
\omega_5(A) = \frac{1}{24 \pi^2} \Tr[ A(dA)^2 + \frac{3}{2} A^3 dA + 
\frac{3}{5} A^5 ] 
\nn\\
&&
\eea
satisfy the descent equations
$d \omega_5$ $=$ ${1\over 24 \pi^2} \Tr F^3$, and
$ \d_v \omega_5 = d \omega^1_4$
with $F=dA+A^2$, $A=A^a T^a$ and $v=v^a T^a$ as usual, the $T^a$ 
being the generators of ${\rm SU}(4)$ in the fundamental ${\bf 4}$ 
representation. For later use we note that for $T^a$ in a general 
representation $\R$ of ${\rm SU}(4)$, the corresponding quantities 
with the trace taken in $\R$ are
\be
{\omega^1_{2n}}^\R = A(\R)\ \omega^1_{2n}\ , \quad\quad
\omega_{2n+1}^\R =A(\R)\ \omega_{2n+1} ,
\ee
where $A(\R)$ is the anomaly coefficient defined by the ratio of the
$d$-symbols taken in the representation $\R$ and in the fundamental
representation. In general $2n$ or $2n+1$ dimensions, since 
the $d$-symbol is given by a symmetrized trace of $n+1$ Lie algebra 
generators,  it is easy to show that the complex conjugate
representation ${\bf R}^*$ has an anomaly coefficient
\be \label{RR*}
A({\bf R}^*) = (-1)^{n+1} A({\bf R}). 
\ee

Due to the connection of anomalies and Chern-Simons actions in one 
higher dimension, it is natural to expect that the four-dimensional
field theory anomaly is dual to a Chern-Simons action in the gauged 
$AdS_5$ supergravity. This is indeed the case as
was first pointed out in \cite{c3}.   
The tree level supergravity action on $AdS_5$ contains the following
terms \cite{c3,dzf,GRW1,vN1}
\be
\label{sugraaction}
S_{\rm cl}[A]= \frac{1}{4 g^2_{SG}} \int d^5 x  \sqrt{g}\, 
 F_{\mu\nu}^a F^{\mu\nu a}  +
k \int_{AdS_5} \omega_5.
\ee
Note that here $F$ is the field strength associated with the 
five-dimensional gauge field $\tilde A_\m$.
The exact values of the coefficients ${1\over 4 g^2_{SG}}$ and $k$ 
will be important for us. Their ratio is fixed by supersymmetry 
\cite{GRW1,vN1}. They may be obtained by dimensional reduction of 
the ten-dimensional IIB supergravity on $S^5$ using the fact that 
the radius of $S^5$ is given by eq. \eq{uv} as 
$R^4/\ap^2=4\pi g_s N$. Then it is easy to determine the 
normalization of the gauge kinetic energy term and one  finds 
\be \label{gkstd}
g_{SG}^2 = \frac{16 \pi^2}{N^2}, \quad k = N^2.
\ee
Note that the action \eq{sugraaction} with the values \eq{gkstd} 
has been used to compute the 2-point and 3-point correlators of 
the currents $J^a_\m$ in the SYM theory \cite{dzf}. To leading 
order in $N$ this gives the correct result.

In usual considerations of supergravity on $AdS$, one considers 
gauge configurations $\tilde A_\m$ that vanish at the boundary and 
so the Chern-Simons term is gauge invariant since 
$\delta \omega_5= d \omega_4^1$ and the integral vanishes. For the 
considerations of the AdS/CFT correspondence however, we precisely 
want nonvanishing boundary values for $\tilde A_\m$ as explained 
above, cf eq \eq{uxii}. Then under a gauge variation 
$\d_v \tilde A$, the variation of the 
Chern-Simons term is a boundary term 
\be \label{dS}
\d_v S_{cl} = k \int_{S^4} \omega^1_4(v,A) \ .
\ee
(We take the SYM theory to be defined on compactified Euclidean 
space, i.e. on $S^4$.) Now by eq \eq{uxiv}, approximating 
$S_{\rm string}^{\rm eff} \to S_{\rm cl}$ and using \eq{dS} one can read off the ${\rm SU}(4)_R$ anomaly obtained from the supergravity
action \eq{sugraaction}. It is
\be \label{GRanom}
\d_v \Gamma[A]= \d_v S_{\rm cl} 
=  N^2  \; \int_{S^4} \omega^1_4 (v,A)\ ,
\ee
which agrees with the gauge theory computation \eq{YManom} to
leading order in $N$. 

We thus see that the  IIB supergravity action contains a
Chern-Simons term at tree level which can account for the chiral 
anomaly of the gauge theory to leading order in $N$. But there is 
also a mismatch of ``-1'' which is of order $1/N^2$ relative to 
the leading term. As discussed above, this should correspond to a  
1-loop effect in IIB string theory. Thus we are lead to examine 
the string one-loop effective action.

\section{One-loop induced Chern-Simons action}

Loop effects in $AdS_5$ supergravity are technically very diffciult
to compute due to the complicated propagators in $AdS$ geometry. 
Here however, this is possible due to the topological character of 
the Chern-Simons action.

\underline{Fermionic contributions}
 
Consider a Dirac fermion $\psi$ in odd 
dimensions (flat) minimally
coupled  to vector 
bosons $A_\mu$ of a group $G$. 
At the quantum level, a
regularization needs to be introduced to make sense of the theory and
one cannot preserve both the gauge symmetry (small and large) and the
parity at the same time \cite{fcs1,fcs2}.
If one chooses to preserve the gauge symmetry by doing a
Pauli-Villars regularization,  then there will be an induced
Chern-Simons term generated at one loop. The result is independent of
the fermion mass. 
In our notation, the
induced Chern-Simons term is 
\be \label{ind}
\Delta \Gamma = \pm \frac{1}{2}\int \omega^\R_{2n+1} = 
\pm \frac{1}{2} A(\R)  \int \omega_{2n+1},
\ee
where $\R$ is the representation of the Dirac 
fermion. The $\pm$ sign depends on
the regularization and can often be fixed within  
a specific context. 

This result was originally \cite{fcs1,fcs2} obtained  
for fermions coupled to gauge fields in a flat spacetime and has been
extended to full generality for arbitrary curved backgrounds and any
odd dimension $d=2n+1$
\cite{AGPM}. The induced parity violating terms are
given (up to a normalization factor) 
by the secondary characteristic class 
$Q(A,\omega)$ satisfying
\be \label{Q}
dQ(A,\omega) = {\hat A(R)} ch(F) |_{2n+2},
\ee
where $\omega$ is the gravitational connection. Since 
$\hat A(R)=1 +{\cal O}(R^2)$ and $\Tr F=0$ for SU-groups, it is 
clear that for $n=2$ ($d=5$) there are no mixed gauge/gravitational 
terms. 
Also, there can be no purely gravitational term since it would 
correspond to a gravitational anomaly in four dimensions which is 
not possible. Hence for the present case of ${\rm SU}(4)$ with 
$n=2$, \eq{Q} simply reduces to $dQ=ch(F)\vert_6$ giving rise to 
the Chern-Simons action upon descent, which does not depend on the 
geometry of $AdS_5$ at all! Hence the result of \eq{ind} for a Dirac 
fermion in flat space(-time) remains valid on $AdS_5$.

Now we need the particle spectrum of the type IIB string theory on
\ads. The only explicitly known states are the KK states
coming from the compactification of the 10 dimensional IIB 
supergravity multiplet \cite{vN2}. So we will examine them first. 
We will argue in the discussion section that the 
other string states are not likely to modify the result.

Particles in $AdS_5$ are classified by unitary irreducible
representations of ${\rm SO}(2,4)$. ${\rm SO}(2,4)$ has the maximal 
compact
subgroup ${\rm SO}(2)\times {\rm SU}(2)$ $\times {\rm SU}(2) $ 
and so its 
irreducible representations are  labelled by the quantum numbers 
$(E_0, J_1, J_2)$. 
The complete  KK spectrum of the IIB supergravity on \ads
was obtained in  \cite{vN2,GRW2} together with information on  
the representation content under 
${\rm SU}(4)_R$. We reproduce these results in the  table below.
Actually, all fermions are symplectic Majorana, giving half the 
anomaly of a Dirac fermion. But there also is a mirror table with 
conjugate ${\rm SU}(4)_R$ representations and 
${\rm SU}(2) \times {\rm SU}(2)$ quantum numbers exchanged (opposite 
chiralities). So these ``mirror" fermions give the same anomaly as 
those in the table and the net effect is that we may restrict ourselves 
to the fermions of the table treating them as if they were Dirac 
fermions.
\be \label{table}
\begin{array}{lcllr}
         & {\rm SU}(2)\times {\rm SU}(2) & & {\rm SU}(4)_R \\ 
\psi_\mu & (1,1/2) 	&&{\bf 4, 20, \cdots} 
&\leftarrow \\
	 & (1,1/2)  	&&{\bf 4^*, 20^*, \cdots} \\ 
\l       & (1/2,0)   	&& {\bf \;\;\;\, 20^*, \cdots}
&\leftarrow \\
         & (1/2,0)   	&&{\bf 4, 20, \cdots} \\
\l'      & (1/2,0)  	&& {\bf 4^*,20^*,\cdots}
&\leftarrow \\
         & (1/2,0)    	&& {\bf 4, 20, \cdots}\\
\l''     & (1/2,0)   	&& {\bf 36, 140, \cdots}\\
	 & (1/2,0)     	&& {\bf 36^*, 140^*, \cdots}\\
\end{array}
\ee

Notice that the fermion towers always come in pairs with conjugate 
representation content, except for a missing ${\bf 4^*}$ state in 
the first  tower of $\l$. As a result \cite{bc}, all contributions
cancel two by two except for the contribution from the unpaired 
${\bf 4}$ of the $\l$ tower. The net resulting induced Chern Simons 
action is
\be
\Delta \Gamma =
- \frac{1}{2} \int_{AdS_5} \omega_{5}.
\ee
While this is almost what we want, it is only half of the desired 
result. However this is not the whole story. 

\underline{Doubleton multiplet}

There are similar ``missing states'' in the boso\-nic towers. 
Together they are identified with the doubleton multiplet 
of ${\rm SU}(2,2|4)$ which consists of a gauge potential, six scalars and
four complex spinors. These are nonpropagating modes in the bulk of 
$AdS_5$ and can be gauged away completely \cite{vN2,GNW}, which is 
the reason why they don't appear in the physical spectrum. 
These modes are  exactly dual to the ${\rm U}(1)$ factor of the 
${\rm U}(N)$ 
SYM living on the boundary \cite{rev}. We will now show that the 
other half of the induced Chern-Simons action is due to the 
corresponding Faddeev-Popov ghosts.

Let us recall that the doubleton multiplet is absent because it has been
gauged away \cite{vN2} by imposing the gravitino gauge fixing
condition (see also \cite{GNW} for the gauging in the 
case of $AdS_7 \times S^4$ case). The basic idea is that upon
compactifying on $S^5$, the original  supersymmetyries 
in 10 dimensions decompose into an infinite tower of (unwanted) 
supersymmetries according to the Fourier expansion.  This can be 
fixed however by imposing a certain 
condition on the variation 
\be \label{psivar}
\d \psi_\a = D_\a \e + {i\over 2 R} \gamma_\a \e
\ee 
of the gravitino. 
Denote the local coordinates of $AdS \times S^5$ by 
$(x^\mu, y^\a)$. A general spinor $\e$ has the decomposition
\be \label{esum}
\e = \sum \e^{I, \pm} (x)\ \Xi^{I,\pm} (y)
\ee
where $\Xi^{I,\pm} (y)$ are the spinor spherical harmonics on $S^5$ and
satisfy ($\Ds_y=\gamma^\a D_\a, \a=5,\ldots 9$)
\be \label{XiDE}
\Ds_y \Xi^{I,\pm} = \mp i (k+ \frac{5}{2})\frac{1}{R}\Xi^{I,\pm}
\ee
where $k = I \ge 0$  and $\Xi^{I,\pm} $ can be written in terms of
the  killing spinors $\eta^{I,\pm}$ on $S^5$.  
Subsituting \eq{esum} into \eq{psivar} and using \eq{XiDE} we get
\be
\d (\c^\a \psi_\a) = {i\over R}\sum \left[ \mp (k+{5\over 2}) + {5\over 2} \right] \e^{I, \pm} (x)\ \Xi^{I,\pm} (y) \ .  
\ee
So one finds that only the component corresponding to $\Xi^{0,+}$ 
is gauge invariant while all other  components of $\c^a \psi_\a$ 
can be gauged away. 
Thus we arrive at the gravitino gauge fixing condition
\be \label{ggf}
\c^\a \psi_\a (x,y) \sim  \chi^{I_0} (x) \eta^{I_0,+} (y) 
\ee
where $\chi^{I_0} (x)$
are some arbitrary spacetime spinor fields. 
We refer the reader to \cite{vN2} for more details.
Therefore we see that \eq{ggf} is the closest one can get to the gauge
condition $\c^\a \psi_\a =0$. One can also rewrite this condition as 
\be
\psi_\a = \psi_{(\a)} + \chi^{I_0}(x) \c^\a \eta^{I_0,+}(y)
\ee
where the part $\psi_{(\a)}$ satisfies $\c^\a \psi_{(\a)} =0$. 
The other Killing spinors $\eta^-$ have been gauged away. The
coefficient of $\eta^-$ would be a field in the 
{\bf $4^*$} of ${\rm SU}(4)$ and is precisely the doubleton 
spinors we are after. 
Now the constraint can be taken care of in 
the functional approach by  introducing in the path
integral  the factor
\bea
&&\int db d \bar{b}
  \frac{1}{\det M} e^{\int d^5 x \bar{b} M b} \\
&&\delta(\c^\a \psi_\a - \sum \chi^I \eta^{I,+} -b^I(x) \eta^{I,-}(y) )
\cdot \delta( \mbox{h.c.}) \nn
\eea
where $b(x)$ is a complex fermionic field 
in the {\bf $4^*$} of ${\rm SU}(4)$  
and $M = \Ds_x$. 
Integrating over $b, \bar{b}$ results in
the gauged fixed lagrangian. The factor $(\det M)^{-1}$ can be handled
by intoducing ghost fields $c, \bar{c}$, which are
bosonic spinor fields on $AdS_5$ and are 
in  the {\bf $4^*$} of ${\rm SU}(4)$. Thus
\be
 \frac{1}{\det M} = \int dc d\bar{c} \; e^{-\int  d^5 x\bar{c} M c}.
\ee
and so give rises to another -1/2 contribution to the induced
Chern-Simons action. 
So altogether we get a total induced Chern-Simons term of -1, 
\be
\Delta \Gamma =
-  \int_{AdS_5} \omega_{5},
\ee 
which is exactly the desired result. 
Notice that the induced
Chern-Simons action (coming with a constant integer coefficient) 
is independent of the radius $R$ 
and this is consistent with the AdS/CFT proposal
since the anomaly and its corrections are independent of $\l$. 

\underline{Bosonic contributions}

There is another interesting effect related to the Chern-Simons action. 
It is known that in three dimensions, the gluons at one loop 
can modify the coefficient of the  Chern-Simons action by an integer 
shift. It has been  argued that \cite{c3,bc}  
there is no such shift for the present case. 
Therefore only spinor loops contribute to the induced Chern-Simons
action and  we find that at finite $N$, the
coefficient $k$ is shifted by
\be \label{kshift}
k \rightarrow k-1\quad\mbox{or}\quad N^2 \rightarrow N^2 -1
\ee
due to the quantum effects of the full set of Kaluza-Klein states. 

A few comments about the absence of a shift due to gluon loops 
are in order. 
The bosonic shift in pure 
Chern-Simons theory was first computed in \cite{wit} 
using a saddle point approximation.  
Later calculations trying to reproduce this results from
the perturbative point of view revealed that the 
precise shift depends on the choice of 
regularization scheme
\footnote{
We thank R. Stora for a useful discussion about the issues of
regularization. 
}.
In the present case of 5-dimensions, 
one may try to employ a regularization scheme and do a
1-loop perturbative calculation to determine the possible shift. 
However, like in the 3-dimensional case, 
it  can be expected that the result
will depend on the choice of regularization and a better way
to determine the shift is called for. One possibility might be to do a
string theory calculation by embedding the Chern-Simons action in a
string setting and to determine the quantum loop effects from the 
string loop effects. Since string theory is free from divergences, 
no regularization related
ambuigities should occur and a definite answer can be expected.

\section{Discussion}

We have reproduced the correct shift $\ N^2\ \to\ $ $N^2-1$ of the 
anomaly 
coefficient as a one-loop effect in IIB supergravity/string theory 
on $AdS_5\times S^5$. This shift is entirely due to the towers of 
Kaluza-Klein states. No massive string states need to be invoked. 
It is indeed likely that the latter play no role at all since 
anomalies are usually due to massless fields only. Note however 
that we need  the full towers of Kaluza-Klein states to get the 
correct result. A truncation to five-dimensional $AdS$ supergravity 
alone would not give the desired result. Also the $AdS_5$ 
supergravity Chern-Simons term originates from compactifying the 
full IIB supergravity. This is another indication that string states 
beyond the Kaluza-Klein towers are unlikely to modify our result. 

We have been able to obtain a non-trivial one-loop result within a 
particularly favourable case. In general, one-loop calculations in 
$AdS_5$ are very difficult - already tree computations are quite 
non-trivial! Of course, the anomaly coefficient $\sim N^2-1$ should 
be exact and there cannot be any further higher-loop corrections 
$\sim N^2 {1\over N^4}$. Again, since the induced Chern-Simons term in 
5 dimensions is closely related to anomalies in 4 dimensions, we 
expect some sort of non-renormalisation theorem to be at work, 
although it would be nice to have a proof of this statement.

Finally, we would like to make some comments on more or less related 
situations. There are other dualities like those involving 
$AdS_7\times S^4$ where one can expect similar Chern-Simons terms 
and doubleton multiplets. The issue of the trace-anomaly in $AdS_5$ 
should be closely related to the present study. Already the leading 
$N$ behaviour of this conformal anomaly is non-trivial to establish 
\cite{a1} and to explicitly obtain the subleading terms might well 
turn out to be more difficult than for the chiral anomaly studied 
here. Effects that are of lower order than $N^2$ have also been 
considered in \cite{a2} which essentially studies situations where 
the leading effect corresponds to open strings at tree level and 
hence comes with just one power of $N$. A somewhat related discussion 
is \cite{a3}. 

It will also be interesting to investigate these anomaly 
issues within the non-commutative version of the AdS/CFT 
correspondence \cite{ncads} to see the origin of the higher 
derivative Chern-Simons terms \cite{chu} on the supergravity side.

\acknowledgments

This work was partially supported by the Swiss National Science
Foundation, by the European Union under TMR contract
ERBFMRX-CT96-0045.


\ed